\documentclass[twocolumn,showpacs,aps,floatfix,pra,superscriptaddress]{revtex4}
\usepackage{graphicx}
\usepackage{bm}
\usepackage{amsmath}
\newcommand{\ignore}[1]{}

\def\beq{\begin{equation}}
\def\eeq{\end{equation}}
\def\la{\langle}
\def\ra{\rangle}
\def\wh{\widehat}
\def\beqa{\begin{eqnarray}}
\def\eeqa{\end{eqnarray}}
\begin{document}
\title{Disclosing hidden information in the quantum Zeno effect:\\  
Pulsed measurement of the quantum time of arrival} 
\author{J. Echanobe}
\email{javi@we.lc.ehu.es}
\affiliation{Departamento de Electricidad y Electr\'onica, 
UPV-EHU, Apartado 644, 48080 Bilbao, Spain}
\author{A. del Campo}
\email{adolfo.delcampo@ehu.es}
\author{J. G. Muga}
\email{jg.muga@ehu.es}
\affiliation{
Departamento de Qu\'\i mica-F\'\i sica, UPV-EHU, 
Apartado 644, 48080 Bilbao, Spain}
%   
%\author{L. Schulman}
%\email{dwsprung@mcmaster.ca}
%\affiliation{
%New York\\
%  Hamilton, Ontario L8S 4M1 Canada}
%\maketitle
%
\begin{abstract}
Repeated measurements of a quantum 
particle to check its presence in a region of space   
was proposed long ago (G. R. Allcock, Ann. Phys. {\bf 53}, 286 (1969))
as a natural way to determine the
distribution of times of arrival at the orthogonal subspace, 
but the method was discarded because of the quantum Zeno effect:
in the limit of very frequent measurements the wave function is reflected and remains
in the original subspace. We show that by normalizing 
the small bits of arriving (removed) norm, an ideal time 
distribution emerges in correspondence with a classical local-kinetic-energy
distribution. 
% but not with a time-of-arrival distribution.      
\end{abstract}
\maketitle
%
%
%*********************************************************************
%
\section{Introduction}
The theoretical treatment of time observables is one of the 
important loose ends of quantum theory. Among them the time of arrival (TOA) 
has received much attention lately
\cite{dam02,rus02,alo02,ON03,gal04,rus04,
bon04,gdme05,LL,as06,heg06,gas06,heg07,man07,tor07,goz07}, 
for earlier reviews see \cite{ML00,MSE02}. 
A major challenge is to 
find the connection between {\it ideal} TOA distributions,
defined for the particle in isolation, formally independent of the 
measurement method, and {\it operational}
ones, explicitly dependent on specific measurement models and procedures.
It is important to know, for example, what exactly a given operational procedure is measuring, 
or if and how a given ideal quantity may or may not be obtained with 
a particular experiment. 

\begin{figure}
%\epsfysize=7.6cm
%\centerline{\epsfbox{Zeno.eps}}
\includegraphics[width=6cm]{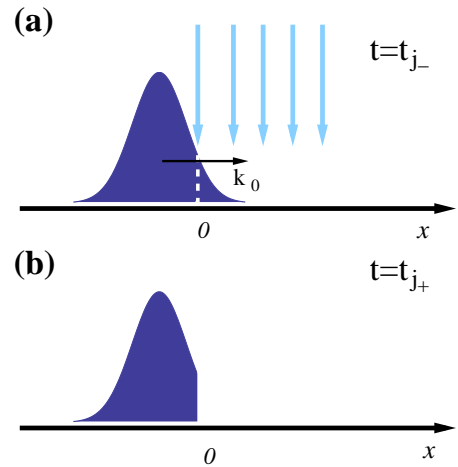}
\caption[]{Schematic representation of the time of arrival measurement by 
periodic projection of the wave function onto the subspace $x<0$
at times $t_j$, $j=..,-1,0,1,..$ separated by $\delta t$.  
Figures (a) and (b) represent two instants immediately before
and after the elimination of norm at $x>0$.   
\label{f0}}
\end{figure}

%In this article we continue this investigation by making further
%connections between operational and ideal quantities, in particular, 
%we shall provide the relation between continuous and pulsed procedures 
%which have been proposed to  measure the arrival time.   
%
Modern research on the quantum TOA owes much to the 
seminal work by Allcock \cite{Allcock}.      
Looking for an ideal quantum arrival-time concept, 
he considered that arrival-time
measuring devices
should rapidly transfer any probability that appears at $x>0$
($x=0$ being the ``arrival position'')
from the incident channel into various
orthogonal and time-labeled measurement channels.
As a simple model
to realize this basic feature he proposed a pulsed, periodic removal,
at time intervals 
$\delta t$, of the wave function for $x>0$, while the $x<0$
region would not be affected, see Fig. \ref{f0}. 
A similar particle removal would provide the 
distribution of first arrivals   
%at a certain region in $\delta t$ 
for an ensemble of classical, freely-moving particles as $\delta t\to 0$.
  
The difficulties to solve the corresponding mathematical 
problem lead Allcock to study instead a different, continuous model with an 
absorbing imaginary potential in the right half-line, $-iV_0\Theta(x)$,
$V_0>0$, 
to simulate the detection.
He argued that the two models should lead to
similar conclusions with a time resolution of the order   
$\delta t$ in the chopping model 
or $\hbar/2V_0$ in the complex potential model.    
He then solved the Schr\"odinger equation with the complex potential 
and noticed that for $V_0\gg|E|_{max}$,
where $E_{max}$ is a maximum 
relevant energy in the wave packet, 
%------------------------------------------------------
%-----------------------------------------------------------
%\footnote{He considered ``source'' boundary 
%conditions instead of standard initial value boundary conditions,
%so that negative energies were allowed in the formalism,
%but the negative energy contribution  
%vanishes if the source is far away.
%His analysis may be repeated with standard initial value 
%conditions without altering the results.})
the apparatus
response vanishes, $-\delta N/\delta t\to 0$, with $N=\la \psi|\psi\ra$, 
because of quantum mechanical reflection. This is one of the first 
discussions of the quantum Zeno effect, although it was not known by
this name at that time. Eight years later Misra and Sudarshan 
\cite{MS77} generalized this result studying the passage of a system
from one predetermined subspace to its orthogonal subspace:
the periodic projection method in the limit $\delta t\to 0$  
was presented as a natural way to model a continuous measurement, 
however it did not lead to a time distribution of the passage but to its 
suppression \cite{MS77}. 
This lack of a ``trustworthy algorithm'' 
to compute TOA and related distributions prompted them 
to put in doubt the completeness of quantum theory and
has been much debated since then, see 
\cite{hom97,fac01,kos05} for review.  
For the specific case of a projection onto a region of space, as in the
TOA procedure proposed by Allcock, several works have
later emphasized 
%with various degrees of mathematical rigour 
that, in the limit of infinitely frequent measurements, 
the resulting
``Zeno dynamics'' in the original subspace corresponds to hard wall (Dirichlet) boundary
conditions \cite{fac01a,fac04,exn05}. 
  
Allcock discarded the short $\delta t$ (or high $V_0$) limit as useless, 
and examined the other limit where the measurement may
be expected to be $V_0$-independent, $V_0\ll|E_{min}|$, later on. 
We shall not treat this limit hereafter but, for completeness, let us 
briefly recall that he got the current density $J(t)$ by deconvolution of the 
absorbed norm with the apparatus response function, 
\begin{equation}
  J(t) = \frac{\hbar}{2m} \la{\psi_f(t)}| (\widehat{k}\delta(\widehat{x}) +
  \delta(\widehat{x})\widehat{k}) |{\psi_f(t)}\ra,
\end{equation}
where $\wh{k}$ and $\wh{x}$ are the wavenumber and position operators, $m$ 
is the mass, and 
the average is computed with the freely moving wave function
$\psi_f$.
This result makes sense classically, 
but in quantum mechanics $J(t)$ is not positive semidefinite even for 
states composed only by positive momenta \cite{Allcock,ML00}.
He also got, within some 
approximation,  a positive distribution,
$\Pi_K(t)$, which has been later known as ``Kijowski's distribution''
\cite{Kijowski74,ML00,ON03},
\begin{equation} \label{eq:Kijowski}
\Pi_K(t) = \frac{\hbar}{m} \la{\psi_f(t)}| \widehat{k}^{1/2}
\delta(\widehat{x}) \widehat{k}^{1/2} |{\psi_f(t)}\ra. 
\end{equation}
In this paper, instead of disregarding the  
Zeno limit because of
reflection, as it is customary, we shall 
look  at the small amount of norm $\delta N$ detected, i.e., eliminated 
in the projection process, and normalize, 
\beq
\label{PiZeno}
\Pi_{Zeno}=\lim_{\delta t\to 0}\frac{-\delta N/\delta t}{1-N(\infty)}.
\eeq 
In this manner, a rather simple ideal quantity will emerge:   
%(It turns out that the method does not really provide a time-of-arrival distribution 
%but a local-kinetic energy distribution.) 
there is, in other words,  interesting
physical information hidden behind the Zeno effect,
which can be disclosed by normalization.  
%Another recent positive use of the Zeno 
%effect has been described recently \cite{NTY}, see also Ketterle \cite{ket07}. 
%When this is done, it is possible to extract two different ideal quantities,
%$J$ and $\Pi_{Zeno}$ in the two limits considered by Allcock
%(strong and weak $V_0$). 
%In parallel, it is possible to obtain these quantities in different 
%parameter regimes of the atom-laser fluorescence experiment. 
%
To fulfill this program we shall put the parallelism 
hinted by Allcock between the pulsed measurement
and the continuous measurement on a firmer, more quantitative 
basis.   

%removal and the 
%In other words, we shall first show the relation 
%between the pulsed procedure and the continuous one. Since the continuous method 
%is mathematically simpler, we shall thus be able to get the physical meaning of the 
%periodic removal process in the fast chopping limit.         
%in the limit of a strong absorbing potential. 
%Because of the existence of the mentioned regime where the atom-laser complex
%potential models coincide, we are actually providing an especific 
%physical content to the use of complex potentials.  

\section{Zeno time distribution}
%
%{\it Zeno time distribution.}
%
We shall now define formally the pulsed and continuous measurement models
mentioned above and also an intermediate auxiliary model
\cite{alo02} that will be a useful bridge between the two.
Ref. \cite{mug06} is followed initially  
although the analysis and conclusions will be quite different.     

%{\it{The pulsed model.}}   
The ``chopping  process'' amounts to a periodic projection of the wave
function onto the 
$x<0$ region at instants separated by a time interval $\delta t$. 
There is nothing here beyond standard measurement theory \cite{vN}. Each chopping step 
eliminates interference terms in the density operator between right and left components,
and the right component is separated from the ensamble (detected) 
so that it cannot come back to the left.  
The wave functions 
immediately after and before 
the projection at the instant $t_j$, are related by     
%
%
%\beq
$
\psi(x,{t_j}_+)=\psi(x,{t_j}_-)\Theta(-x). 
$
%\eeq
%

%{\it{The kicked model.}}
%
The wave at $x>0$ may also be canceled with a
``kicked'' imaginary potential  
%
%\beq
%\label{kick}
$
\wh{V}_k=\wh{V}\delta t\, F_{\delta t}(t),
$
%\eeq
%
where the subscript ``k'' stands for ``kicked'' and 
%
%\beqa
$F_{\delta t}(t)=\sum_{j=-\infty}^{\infty} \delta(t-j\delta t)
$,
\beq
\wh{V}=-i\wh{V}_I=-iV_0\Theta(\wh{x}),
\label{v}
\eeq
provided  
\beq
\label{V0t}
V_0\delta t \gg \hbar.
\eeq
The general (and exact) evolution operator 
is obtained by repetition of the basic unit
\beq\label{ue}
\wh{U}_k(0,\delta t)=
e^{-i \wh{H}_0\delta t/\hbar}e^{-i\wh{V}\delta t/\hbar},
\eeq
where $\wh{H}_0=-(\hbar^2/2m)\partial^2/\partial x^2$.
%is the kinetic energy operator.  
%The reason for having introduced  $\delta t$ (and no other quantity with
%dimensions of time)
%in Eq. (\ref{kick}) should now become clear. 

For the continuous model, the evolution 
under the imaginary potential (\ref{v})
is given by  
%
%\beq
%{V}=-iV_0\Theta(x-X_c),
%\eeq
%
\beqa\label{uot}
\wh{U}(0,\delta t)&=&
e^{-i(\wh{H}_0+\wh{V})\delta t/\hbar}=e^{-i\wh{H}_0\delta t/\hbar}
e^{-i\wh{V}\delta t/\hbar}
\nonumber\\
&+&{\cal O}(\delta t^2[\wh{V},\wh{H}_0]/\hbar^2).
\label{condi}
\eeqa
%
%The infinitesimal evolution operators converge,
%and in fact the same happens for  
%an arbitrary  evolution interval by Trotter's formula,   
%
%\beq
%\wh{U}_k(0,t)\to\wh{U}(0,t),\;
%{\rm when} \Delta\to 0,\, {\rm or}\, V\, {\rm fixed},
%\eeq
%
%where $t=N\Delta t$ is kept fixed and finite.
Comparing with Eq. (\ref{ue}) we see that the kicked and continuous models 
agree when 
\beq
\delta t^2|\la[\wh{V},\wh{H}_0]\ra|/\hbar^2 \ll 1.
\label{kc}
\eeq
At first sight a large $V_0\delta t/\hbar$, see Eq. (\ref{V0t}), seems 
to be incompatible with this condition so that the three models 
would not agree \cite{mug06}.   
%:  
%by satisfying one of the inequalities 
%connecting between two models,  
%say kicked and chopping in Eq. (\ref{V0t}),   
%Eq. (\ref{kc} that relate kicked and complex potential would not be 
%satisfied.    
% Thus for chopping intervals below $1/V_0$
%chopping and complex 
%potential models will generally disagree,  
%whereas for chopping intervals above $1/V_0$
%the two complex potential models 
%(kicked and continuous)
%will also disagree.''
In fact the numerical calculations give a different result
and show a better and better agreement between 
the continuous and pulsed models as $V_0\to \infty$
when $\delta t$ and $V_0$ are linked by some predetermined 
(large) constant $\alpha$, 
%  
%\beq
$\delta t=\alpha\hbar/V_0.$ 
%\label{al}
%\eeq
%----------------------------------------------------------------------------
\begin{figure}
%\epsfysize=7.6cm
%\centerline{\epsfbox{Zeno.eps}}
\includegraphics[height=6.4cm]{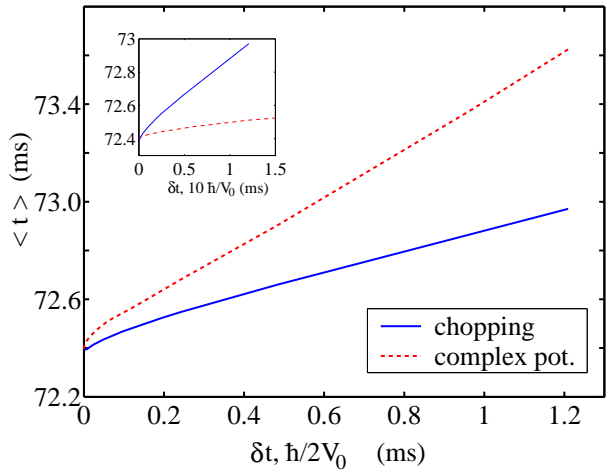}
\caption[]{Average absorption times evaluated from $-dN/dt$ (normalized) 
for the projection method and the continuous (complex potential) model.  
The initial state is a minimum uncertainty product Gaussian for $^{23}$Na atoms 
centered at $x_0=-500\mu$m with $\Delta x=23.5\mu$m and average velocity $0.365$
cm. 
In all numerical examples negative momentum components of the initial state
are negligible.
\label{f1}}
\end{figure}
%--------------------------------------------------------------------------------
\begin{figure}
%\epsfysize=7.6cm
%\centerline{\epsfbox{Zeno.eps}}
\includegraphics[height=6.2cm]{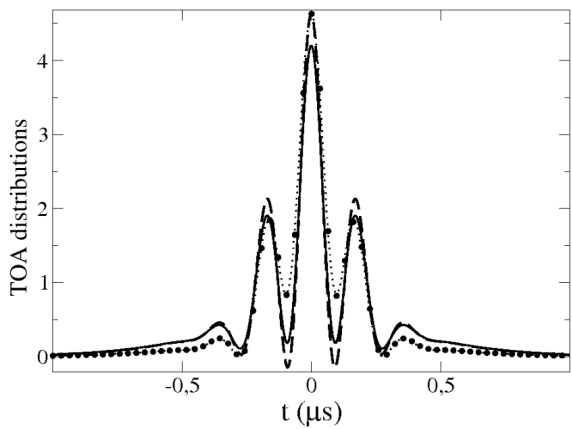}
\caption[]{Time-of-arrival distributions: Flux $J$ (dashed line),
$\Pi_K$ (solid line), 
$\Pi_{Zeno}$ (big sparse dots), and $\Pi_{chopping}$ (for pulses separated by 
$\delta t=0.266$ ns, dotted line).
The initial wave packet is 
a combination $\psi=2^{-1/2}(\psi_1+\psi_2)$
of two Gaussian states for the center-of-mass motion of a single Caesium atom
that become separately minimal uncertainty packets (with
$\Delta x_1=\Delta x_2=0.021\,\mu$m,
and average velocities 
$\la v\ra_1=18.96$ cm/s, 
$\la v \ra_2 =5.42$ cm/s) at $x=0$ and $t=0\,\,\mu$s.
\label{f2}}
\end{figure}
%------------------------------------------------------------------------
\begin{figure}
%\epsfysize=7.6cm
%\centerline{\epsfbox{Zeno.eps}}
\includegraphics[height=6.2cm]{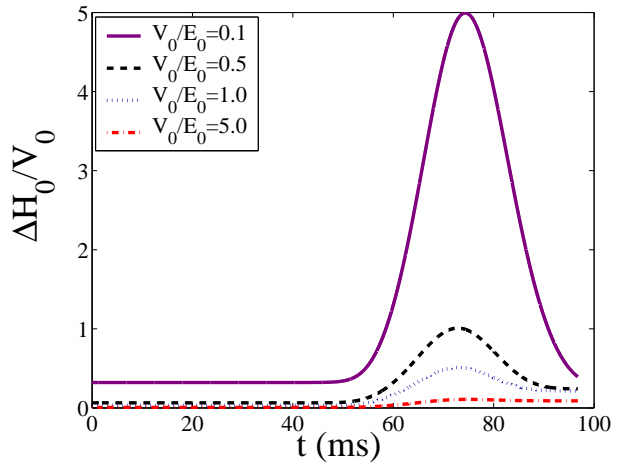}
\caption[]{Diminishing ratio $\Delta H_0/V_0$ with increasing $V_0$ versus time
for the same initial wave function as in Fig. 2 ($^{23}$Na atoms.) 
$E_0$ is the initial kinetic energy.    
\label{f3}}
\end{figure}
%-----------------------------------------------------------------------
%
Figures \ref{f1} and \ref{f2} illustrate this agreement: 
in Fig. \ref{f1} (inset) the average absorption time, 
$\la t\ra=\int^\infty_0 (-dN/dt)t dt/[1-N(\infty)]$, is shown versus $\delta t$ and 
$10\hbar/V_0$ for a Gaussian wave packet sent from the left 
towards the origin. 
%, the edge point 
%where chopping or absorption by $-iV_0\Theta(x)$ takes place.
%Note the convergence of the two straight lines as $\delta t$ 
%and $\hbar/V_0\to 0$.
%Not only the mean values but also the variances 
%(not shown) tend to agree as $\delta t\to 0$.
The main figure shows the time averages versus $\delta t$ (chopping) 
and $\hbar/2V_0$ (continuous). The lines bend at high coupling
because of reflection.     
The normalized absorption distribution as $V_0\to \infty$ 
was derived in 
\cite{ked05} by making explicit use of the known stationary 
scattering wave-functions, 
\begin{equation}
\Pi_N(t) =
\frac{\hbar}{mk_0}\la{\psi_f(t)}|
\widehat{k}\delta(\widehat{x})\widehat{k}|{\psi_f(t)}\ra,
\label{pz}
\end{equation}
where $k_0\hbar$ is the initial average momentum. 
Fig. \ref{f2} shows for a more challenging state, 
a combination of two
Gaussians, 
that this ideal distribution
becomes indistinguable from the normalized chopping distribution 
when $\delta t$ is small enough. Even the minor details are reproduced, and 
differ from $J$ and $\Pi_k$, also shown.   
     
To understand the compatibility ``miracle'' of  
the inequalities (\ref{V0t}) and (\ref{kc}), we apply 
the Robertson-Schr\"odinger
(generalized uncertainty principle) relation,    
\beq
\left|\la [\wh{V},\wh{H}_0]\ra\right|\le 2 |\Delta V_I| \Delta H_0,    
\eeq
where $\Delta$ denotes the standard deviation.  
Since $|\Delta V_I|$ is rigorously bounded at all times by 
$V_0/2$ \footnote{If $N_+$ is the norm in 
$x>0$, $|\Delta V|=V_0(N_+-N_+^2)^{1/2}$ which is maximal at 
$N_+=1/2$.},   
imposing $\delta t V_0=\alpha\hbar$ with $\alpha\gg 1$,
a sufficient condition for dynamical agreement among the models 
is 
\beq
%\frac{V_0}{\Delta H_0}>>\alpha.
V_0\gg\Delta H_0. 
\label{condi3}
\eeq
For large $V_0$ the packet is basically reflected by the wall so that
$\Delta H_0$ tends to retain its initial value and 
Eq. (\ref{condi3}) will be satisfied during the whole propagation.  
Fig. \ref{f3} shows the ratio $\Delta H_0/V_0$ for three values of $V_0$ as
a function of time. 
 
This implies in summary that $\Pi_{Zeno}=\Pi_N$,   
Eq. (\ref{pz}), a very remarkable result, which illustrates  
that an active intervention on the system dynamics may 
after all provide an ideal 
quantity  defined for the system in isolation.

%
%
%
%
%
%
%
%\section{A more general relation between pulsed 
%and continuous measurements}
%
%
%
%
%
%
%
%
%
\section{An approximate relation between pulsed 
and continuous measurements}
So far we have discussed the limits 
$\delta t,\hbar/V_0\to 0$ in order to find the 
corresponding time distribution.    
In fact a very simple argument relates 
the pulsed and continuous measurements {\em{approximately}} for finite, 
non-zero values of $\delta t$ and $\hbar/V_0$, when 
they are sufficiently large to make reflection negligible:   
the average detection time is delayed with respect to the ideal limit 
corresponding to $\Pi_{Zeno}$ as
\beq
\la t\ra \approx \la t\ra_{Zeno}+\delta t/2\approx
\la t\ra_{Zeno}+\frac{\hbar}{2V_0}, 
\eeq
see Fig. \ref{f1}, since, once a particle is in $x>0$, 
$\frac{\hbar}{2V_0}$ and $\delta t/2$ are precisely the average life times 
in the continuous and discrete measuring models, respectively \footnote{The origin ordinate would be slightly above $\la t\ra_{Zeno}$ for 
optimized straight lines. Reflection
at small $\delta t$ (or high $V_0$)
favors the detection of faster particles and bends the 
$\la t\ra$ lines towards shorter times,
as in Fig. \ref{f1}.}. 
This suggests an approximate agreement between projection 
and continuous dynamics provided that the relation 
%
%\beq
$\delta t\approx\hbar/V_0$
%\label{sch}
%\eeq
%
is satisfied. 
For large $V_0$, this is asymptotically not in contradiction with the
requirement of a large $\alpha$ since  
$V_0^{-1}-(\alpha V_0)^{-1}\to 0$ as $V_0\to\infty$; in any case  
quantum reflection breaks down the linear dependence,
see Fig. \ref{f1}.   
    
A similar relation between pulsed and continuous measurements 
was described by Schulman \cite{Schulman98} and has been tested 
experimentally \cite{ket07}. The simplest model in \cite{Schulman98} may be reinterpreted 
as a two-level atom in a resonant laser field, with the excited state   
decaying away from the 2-level subspace at a rate $\gamma$ \cite{mug06}, 
$\widehat{H}=\frac{\hbar}{2}\left({0\atop \Omega} {\Omega\atop{-i\gamma}}\right)$. 
The  relation between pulsed and continuous measurements   
follows by comparing  the exponential decay 
for the effective 2-level Hamiltonian 
with Rabi frequency $\Omega$ and excited state lifetime $1/\gamma$, with the 
decay dynamics when $\gamma=0$ and the system is projected every $\delta t$ into
the ground state.    
It takes the form \cite{Schulman98}
%
%\beq
$\delta t =\frac{4}{\gamma}$
%\label{sr}
%\eeq
%
for $\gamma/\Omega\ll 1$ (weak driving). 
%Note that $1/\gamma$ is the life time for an excited atom, but 
%for an atom starting in the ground state the lifetime in the 2-level subspace 
%is in fact $\gamma/\Omega^2$ because of the slow pumping to the excited state.   
In our TOA model we have a different set of parameters but a comparison
is possible by taking into account that the imaginary potential (\ref{v}) may be
physically interpreted as the effective interaction for the ground state 
in the weak driving regime, for a localized resonant laser excitation 
with subsequent decay, $\widehat{H}=\widehat{H}_0+\frac{\hbar}{2}\left({0\atop \Omega\Theta(\widehat{x})} {\Omega\Theta(\widehat{x})\atop{-i\gamma}}\right)$. 
This gives \cite{ON03,rus04} 
%
%\beq
$V_0=\frac{\hbar\Omega^2}{2\gamma}$,
%\eeq
%
so that $\delta t\approx\hbar/V_0$
%Eq. (\ref{sch})
becomes
\beq
\delta t \approx 2\frac{\gamma}{\hbar\Omega^2},
\eeq
different from Schulman's relation, as it may be expected since 
the pulsed evolution depends on $\Omega$ in Schulman's model 
but not in our case, where it is only driven by the kinetic energy 
Hamiltonian $\wh{H}_0$.         
   
\section{Discussion}
%   
%{\it Discussion.} 
%  
The first discussions of the Zeno effect, understood as 
the hindered passage of the system between orthogonal subspaces because of 
frequent instantaneous measurements, 
emphasized its problematic status and regarded it as a failure 
to simulate or define quantum passage-time distributions. 
We have shown here 
that in fact there is a ``bright side'' of the effect:  
by normalizing the little bits of norm removed at each projection 
step, a physical time distribution defined for the freely moving system
emerges. (There are other 
``positive'' uses of the Zeno effect, such as reduction of decoherence in quantum computing, see e.g. \cite{NTY,f05}.) In the case of the projection measurements to determine 
the TOA, this distribution is given by Eq. (\ref{pz}).    
This result is fundamentally different from
the current density 
or from Kijowski's TOA distribution,  
(operational approaches to measure them by fluorescence are described in  
\cite{dam02, ON03})  
and corresponds to a classical time-distribution of local 
kinetic energy density \cite{coh79,ked05}, rather than 
a classical TOA distribution \footnote{$\Pi_K$ could in principle 
be obtained in the Zeno limit for states with positive momentum support 
by transforming the initial state 
$\la k|\psi_f\ra\to k^{1/2}\la k|\psi_f\ra/C$, with $C$ a constant,   
as in the Operator Normalization technique \cite{ON03}.}.      
 
%The Zeno distribution is 
%thus not in correspondence with a classical time-of-arrival.     

Experimental realizations of repeated measurements will rely on 
projections with a finite frequency and pulse duration 
that provide approximations
to the ideal result. Feasible schemes may be based on 
pulsed localized resonant-laser excitation \cite{ket07}, or 
sweeping with a detuned laser \cite{sweep}.
A challenging open question for further research is the possibility to devise
alternative (non periodic) pulse sequences to enhance the robustness 
of the Zeno effect in a TOA measurement, similar in spirit to the ones proposed in the 
context of quantum information processing \cite{Lidar}. 
Nevertheless, the ordinary (periodic) sequence has successfully been 
applied in quantum optical realizations of the quantum Zeno effect \cite{ket07}.   
    
The proposed normalization method may be applied to other 
measurements as well, i.e. not only for a TOA of 
freely moving particles, but in general 
to first  passages between orthogonal subspaces,  
and it will be interesting to find out in each  
case the ideal time distribution brought out by normalization.     

\begin{acknowledgments}
This work has been supported by Ministerio de Educaci\'on y Ciencia (BFM2003-01003), and UPV-EHU (00039.310-15968/2004). A. C. acknowledges
financial support by the Basque Government (BFI04.479).
\end{acknowledgments}

%\begin{references}

\end{document}